# Unlocking high coercivity at room temperature in phase-modified MoS$_2$

Zainab Chowdhry[1*], Kushal Mazumder[2,3], Praveen Hegde[1], Ajith Jena[4], Vidya Praveen Bhallamudi[1], and Pramoda K. Nayak[1,2*.]

[1]Department of Physics, Indian Institute of Technology Madras, Chennai, India
[2]Center for Nano and Material Sciences, Jain (Deemed-to-be University), Jain Global Campus, Bengaluru, India
[3]Department of Physics, School of Sciences, IILM University, Greater Noida, India.
[4]Indo-Korea Science and Technology Center, Bengaluru, India

E-mail: zainab.chowdhry@physics.iitm.ac.in,

pramoda.nayak@jainuniversity.ac.in,

Two-dimensional (2D) materials showing room-temperature magnetism and high coercivity are desired for combining magnetism with semiconducting properties useful for spintronics. In this work, the magnetic properties of the 1T phase of MoS$_2$ have been studied at room temperature. We observe ferromagnetism with a coercivity of ~0.3 T and a maximum saturation magnetization of 0.26 emu/g at room temperature. This is the highest among coercivities reported so far in 2D magnets at room temperature. MoS$_2$ nanosheets are prepared using a single-step hydrothermal synthesis with a relative 1T-phase reaching up to 77%. We report a correlation between the structural and magnetic characteristics via interplanar distance and coercivity. The increase in interplanar distance is also accompanied by an increase in the 1T phase. Our results pave a useful way of controlling the coercivity and saturation magnetization in a 2D magnet with applications in spintronics and low-power quantum devices.

## 1. Introduction

Recent discoveries of layered materials and particularly magnetism in them have driven investigations to understand their underlying physics and exploring suitable applications.[1–9] Many two-dimensional (2D) magnets identified exhibit intrinsic magnetic behavior only at low temperatures, with a few exceptions, such as $Fe_5GeTe_2$ and $Fe_3GaTe_2$.[10–12] To overcome this temperature limitation, researchers have explored alternatives like phase and defect engineering[13] in non-magnetic materials.

Defects, such as vacancies, can lead to dangling bonds. These, in turn, lead to unpaired electrons that can result in ferromagnetism. These have shown a lot of promise for room-temperature ferromagnetism; an early example of this was ferromagnetism in graphene.[14] The search for magnetic ordering in a semiconducting material at room temperature, which can lead to new device architecture and functionality, has motivated similar studies in other 2D materials. Prime among these is molybdenum disulphide, $MoS_2$. $MoS_2$[15] is a very actively studied 2D material for its diverse applications in electronics,[16–18] optoelectronics,[19] catalysis,[20,21] energy storage,[22] etc. Reports on inducing ferromagnetism in it via doping,[23–25] defect generation,[26–30] edge-aligned growth,[31–33] phase modification,[34,35] strain induced,[29,36] etc., have been present; however, its magnetic properties are not fully explored. The usual 2H phase is diamagnetic, whereas the 1T phase is known to be ferromagnetic.[35,37] Additionally, a good number of studies have discussed the origin of defect-based ferromagnetism in $MoS_2$, such as unpaired Mo ions along zig-zag edges or at the site of sulfur vacancies. A majority of these existing studies focus on the magnetic moment of the sample only, and very few discuss coercivity, which has been limited to only a few tens of Oesterds.[26],[34,37].

Room temperature operation and tunable coercivity are required for any 2D magnet to be integrated into future technologies. Few 2D systems have been reported to show room temperature ferromagnetism with high coercivity, such as 0.15 T in $MnSiTe_3$,[38] 0.08 T in $MnGeTe_3$,[38] 0.14 T in $WSe_2/MoSe_2$[39] heterostructure, and 0.11 T in $CrI_3$.[40] In $MoS_2$, the maximum coercivity reported previously is 0.18 T at 10 K for vanadium-doped $MoS_2$[41] and 0.9T at 100 K for $MoS_2$ codoped with Nb and Co.[24] Hence, it is necessary to explore the magnetic properties of any 2D material at room temperature from the perspective of its coercive field.

Here, we report $MoS_2$ prepared in 1T phase with coercivity reaching 0.3 T at room temperature, which is unreported for a 2D material to the best of our knowledge.[38–40] We explore the influence of structural modification in the lattice on coercivity through strain and magnetic anisotropy analysis. We explore, in particular, the role of interplanar distance and relative 1T phase in controlling the coercivity of a 2D magnet via strain. Our results show an increase in coercivity as the interplanar distance increases, opening the possibility of controlling the anisotropy of lamellar 1T $MoS_2$ nanosheets, which can be a potential 2D magnet.

## 2. Results and Discussions

### 2.1. Structural characterization and phase identification of the synthesized $MoS_2$ samples

We report data on six $MoS_2$ powders, S1-S6, prepared using the hydrothermal method as shown in **Figure 1**. The synthesis details are given in Section 4.2. These are mixtures of 2H and 1T phases of $MoS_2$ as identified by High-Resolution TEM scans on the nanosheets. At larger length scales, we observe the nanosheets' morphology rich in edges. As discussed later, it can play an essential part in defect-driven magnetism.[31–33] The representative data for various characterization techniques: XRD, Raman, and XPS, performed on a sample S1, are presented. The XRD spectrum in Figure 1 (c) shows the characteristic (002) planes of the 1T phase to be

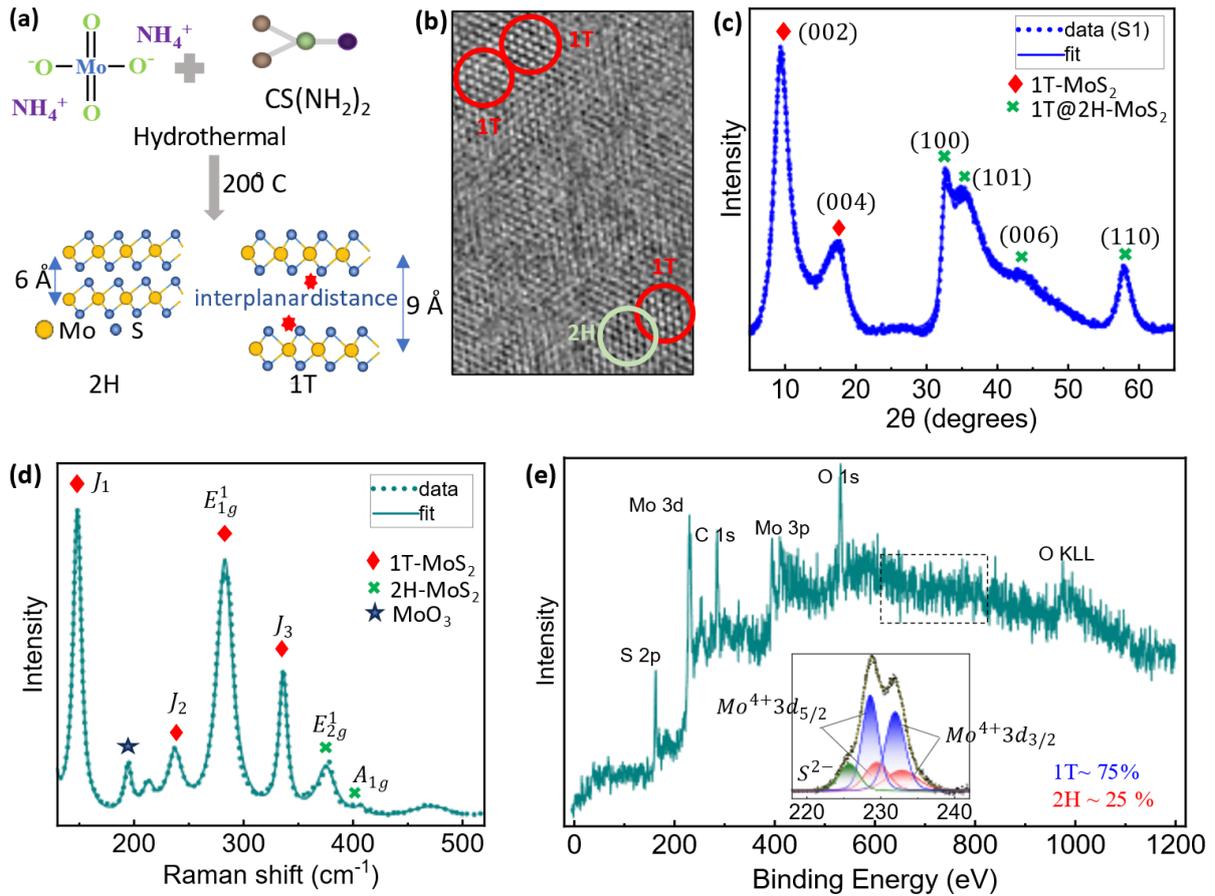

**Figure 1: Structural characterization and phase identification of the synthesized MoS₂ samples** (a) Schematic of the single step hydrothermal procedure which synthesizes MoS$_2$ in 2H and 1T phase, with a larger interplanar spacing in 1T phase. (b) HRTEM images identifying the 1T (octahedral, red) and 2H (trigonal prismatic, green) phases within the synthesized MoS$_2$ nanosheets. (c) XRD scan shows the 1T (red diamonds) and 2H (green cross) phase in the sample. (002) peak for 2H phase expected at ~ 14° is shifted towards 9° characteristic of lattice expansion under 1T phase. (d) Raman spectrum shows the modes characteristic of the 1T phase (e) XPS survey scan show Mo and S as the main ionic species present and absence of magnetic impurities in the sample. (inset) XPS scan around the Mo 3d region shows the deconvolution of the 1T (~ 75%) and 2H phase within it.

at $2\theta$ ~9°. The corresponding peaks from the 2H phase, expected at 14°, are absent. This shows an increase in the interplanar distance from ~6 Å (2H phase, JCPDS Card No. 73-1508)[42] to ~

9 Å. A mixture of peaks in the range of $2\theta = 30° - 65°$ arise from the overlapping peaks of 2H and 1T phases. The atomic arrangement along the basal planes remains the same, and a new lamellar structure gets formed, expanding the $MoS_2$ lattice when the 1T phase is being formed.[42] The (002) orientation is mainly from the 1T phase, supported by the second-order peak from the (004) plane at $2\theta \sim 17°$.

Figure 1 (d) shows the Raman spectrum of the nanosheets. It has the characteristic Raman peaks of 2H $MoS_2$ located at 380 $cm^{-1}$ ($E_{2g}$), 404 $cm^{-1}$ ($A_{1g}$), and 454 $cm^{-1}$ (longitudinal acoustic phonon). The 147, 226, 282, and 335 $cm^{-1}$ peaks correspond to $J_1$, $J_2$, $E_{1g}$, and $J_3$ modes, characteristic of the 1T phase.[35] The $E_{1g}$ band is related to the octahedral coordination of Mo in 1T $MoS_2$. The $E_{2g}$ and $A_{1g}$ modes are associated with the 2H phase and are less intense compared to the $J_1$, $J_2$, $E_{1g}$, and $J_3$ modes characteristic of the 1T phase. The $MoO_3$ peaks observed can be from the laser beam's heat-induced oxidation of $MoS_2$.

The extended survey scan of XPS is presented in Figure 1 (e). It crucially shows the absence of magnetic impurities (black dotted square), besides establishing Mo and S as the main ionic species. XPS scan around the Mo 3d region in the inset shows the deconvolution of the 1T (~ 75%) and 2H phases within it. Details of the measurements, raw data, background subtraction, and other relevant analyses are presented in the **Supporting Information**.

M-H data at room temperature shows a well-defined hysteresis curve, as shown in **Figure 2**. This is a signature of the ferromagnetism coming from the 1T phase. The data is presented after removing the diamagnetic contribution from the background and the 2H phase. Control measurements of the empty cavity, and with just the Teflon tape, have ruled out spurious signals of this level. They are presented along with the raw data in the Supporting Information. Another feature is the large room temperature coercivity $H_c \sim 0.3$ T observed in sample S1. One of the synthesized samples exhibited $H_c$ ~ 0.32 T but is not reported here as its XRD data was not recorded. To our knowledge, this is the highest-reported coercivity in a 2D material at room temperature. There are two noteworthy aspects to this magnetic hysteresis data. We observe a large variability of saturation magnetization and coercivity across the samples, nominally prepared by the same method.

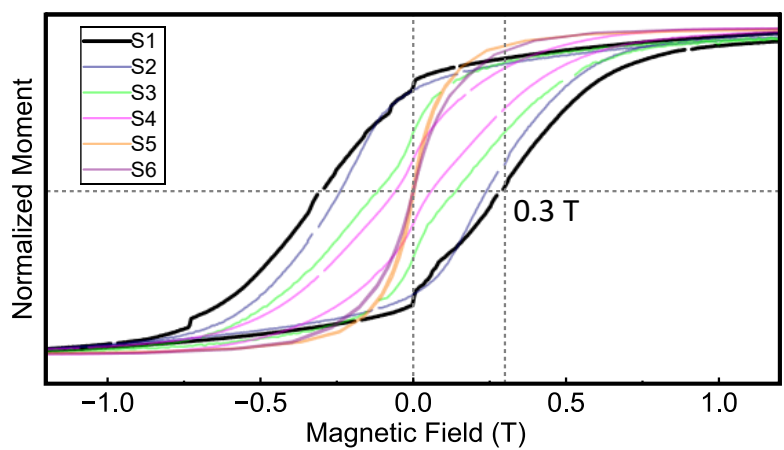

**Figure 2: Observation of high coercivity and room temperature ferromagnetism in synthesized MoS$_2$ samples.** Magnetization vs magnetic field (M-H) measurements taken at room temperature for samples S1-S6. The samples have a varying combination of ferromagnetic hysteresis from the 1T phase and diamagnetic background from the 2H phase. The ferromagnetic data after diamagnetic background subtraction are shown (background details in Supporting Information). The coercivity $H_c$ and saturation magnetization $M_s$ were extracted after removing the diamagnetic background. The data has been normalized in magnetic moment to emphasize the wide hysteresis loops and $H_c$ ~ 0.3 T corresponding to an $M_s$ ~ 0.05 emu/g.

The coercivity $H_c$ and saturation magnetization $M_s$ were extracted after removing the diamagnetic background. The data has been normalized in magnetic moment to emphasize the wide hysteresis loops. In the following sections, the characterization results were extended to study the origin of magnetism, its variability, and the likely causes for the relatively high coercivity seen in our samples.

## 2.2. Origin of magnetism in MoS$_2$

There can be two primary sources of magnetism in MoS$_2$ when there is no external magnetic doping. The spin-resolved band structure of 1T MoS$_2$ calculated by DFT shows spin-polarization at the Fermi level and a calculated moment of $2\mu_B$/(Mo atom). However, most of the reported moments in the 1T phase synthesised from the hydrothermal method are far less than the calculated moment.[34,37] This is also the case for the given samples.

Alternate to the defect-based magnetism from zig-zag edges, folds, and curls which are sites for unpaired Mo ions, Cai[34] presented a combined origin of ferromagnetism from the 1T-MoS$_2$ phase and sulphur vacancies. When the spins of the localized sulphur vacancy align with the spins of the adjacent Mo ions in the 1T phase, an exchange interaction is formed, resulting in enhanced ferromagnetism. To confirm the existence of sulphur vacancies, we performed Electron Paramagnetic Resonance (EPR) spectroscopy at 77 K. **Figure 3** (a) shows a broad asymmetric signal at g=2.029 from unpaired electrons from the sulphur vacancies. The narrow signal at g = 1.92 has been reported from unpaired Mo$^{5+}$ ions.[43–45]

From the Debye-Scherrer equation, the crystal domain size D was calculated for the (002) peak across the samples, and the details of its calculation are provided in the Supporting Information. The crystalline domain size is between 15-21 nm, as shown in Figure 3 (b). The sample with the smallest domain size of D ~15 Å has a larger saturation magnetization of 0.26 emu/g, as this is where the disordered grain boundary or defects in the nanosheets are higher. HRTEM measurements in Figure 3 (c) reveal bends, folds, and curls visible along the edges of the nanosheets, displaying a flowery petal-like structure. These edges are rich in unpaired Mo ions and sulphur vacancies and have been previously predicted to be the origin of edge-based magnetism in $MoS_2$.[31–33] The magnetism seen in our powders likely arises from a combination of the two possible mechanisms mentioned above, with the defect-based mechanism playing a majority role given the measured moments.

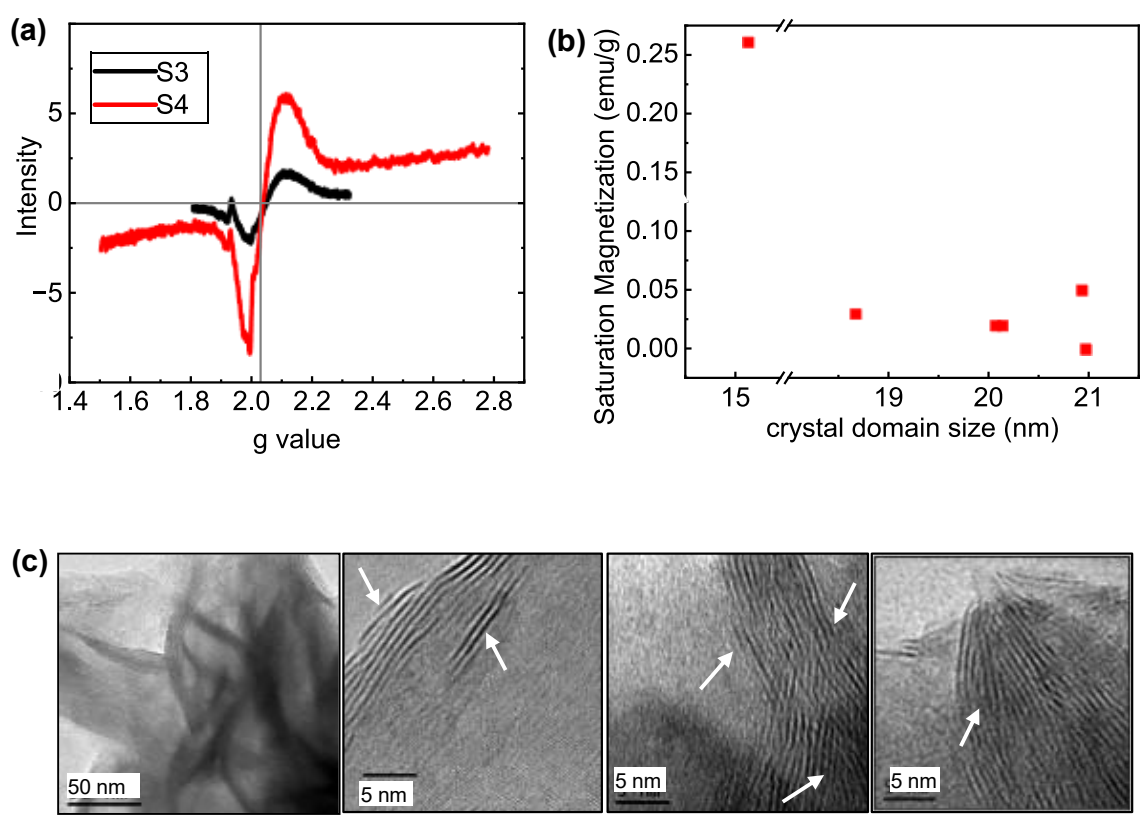

**Figure 3: Presence of vacancies and defects in the MoS₂ sample** (a) EPR measurements at 77 K show a broad asymmetric signal at g=2.029, and a narrow signal at g = 1.92. The g = 2.029 peak can be associated with unpaired electrons from the sulphur vacancy defects. The g = 1.92 peak can be from unpaired $Mo^{5+}$ ions. (b) The crystal domain size D ranged between 15-21 nm. The highest magnetization of 0.26 emu/g has the smallest D ~15 Å. (c) HRTEM measurements reveal a flowery petal-like structure of the nanosheets with bends, folds, and curls (white arrows) along the edges.

## 2.3. Origin of the coercivity variation in MoS₂

Besides the saturation magnetization, the coercivity also varies widely in our samples. We use XRD to justify the collective behavior of the sample, over Raman, which is a more localized study.

**Figure 4** (a) presents the XRD data for all the samples. Figure 4 (b) is a zoomed-in version of (a) around the (002) peak. Also shown are Gaussian fits to this peak (red) after deconvolution of the spectra (details in Supporting Information). From Bragg's law, $2d \sin\theta = n\lambda$, we calculate the interplanar distance 'd' where $2\theta$ is the Bragg reflection angle.

The (002) Bragg reflection angle was observed to vary across the samples from 9.3-9.7 Å. As the (002) peak shifted to lower angles, the interplanar distance increased from 9.1 Å to 9.5 Å. The top panel of Figure 4 (c) presents the coercivity as a function of the interplanar distance. As d increased, the coercivity increased by two orders of magnitude from 0.003 T to ~ 0.3 T. Our results open a route for control of magnetism in a 2D material via its structural properties. The relative 1T phase was calculated as $\text{area }\%_j = \frac{A_j}{\sum_j A_j}$, where $A_j$ is the area under the (002) and (100) reflections in the XRD data corresponding to the 1T-only and the 2H + 1T regions. We observe a systematic increase in 1T content from 32% to 77% with interplanar distance, as presented in the bottom panel. The values extracted from structural and magnetic analysis of the samples are summarized in Table 1 of the Supporting Information.

For our samples, we systematically observe a correlation between the interlayer distance, relative 1T phase %, and magnetic coercivity. The 1T phase % obtained from XPS data matches with the results obtained using XRD, and presented in the Table-1 of Supporting Information The overall results indicate the presence of an intercalated ion in the lattice during synthesis, which, when present in the MoS₂ lattice, can expand it and stabilize the 1T phase. Charge transfer from the intercalated ion to the Mo or S vacancy defects in the lattice energetically favors the 1T phase over the 2H phase, resulting in 1T phase formation.[46]

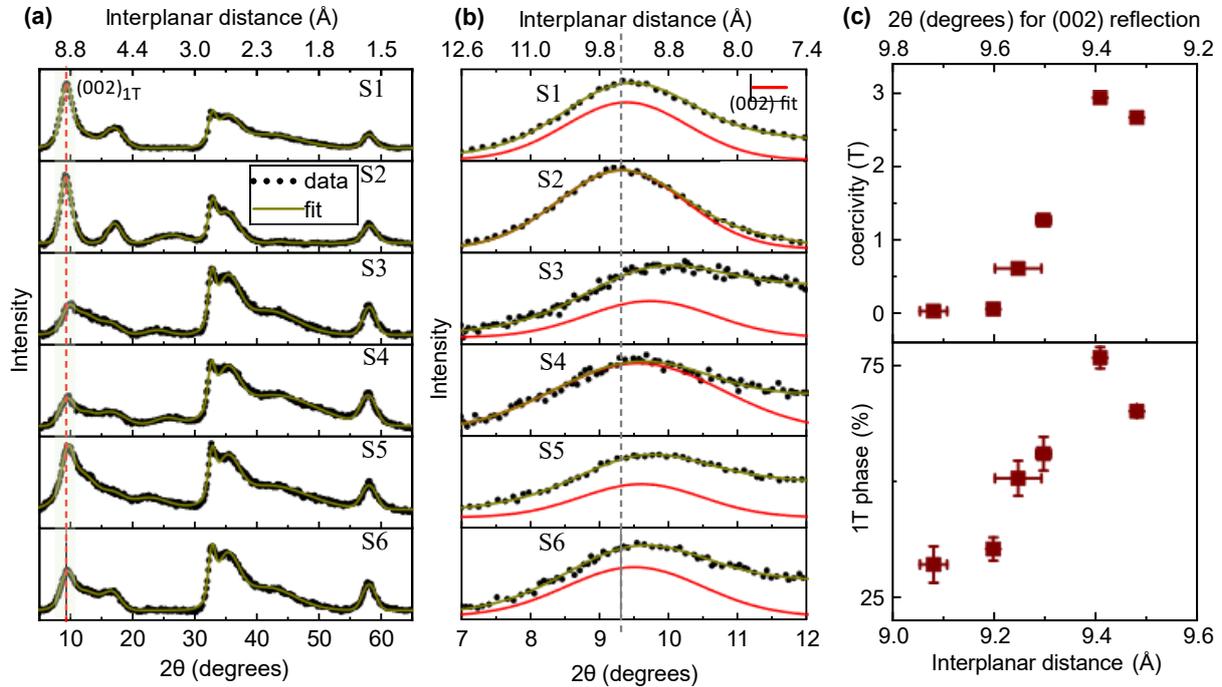

**Figure 4: Dependence of coercivity $H_c$ on the interplanar distance and variation in magnetic properties via structural properties of the MoS$_2$ sample** (a) comparison of XRD data from S1-S6 (b) zoom in around the (002) peak of 1T phase (red) shows a shift towards higher 2θ angles. (c) The top panel shows the coercivity as a function of the interplanar distance. The bottom panel shows the increase in the relative 1T area % from 32 to 77 % as a function of the interplanar distance.

Magnetic anisotropy energy (MAE) plays a crucial role in determining the coercivity of magnetic materials, particularly in 2D systems. Magnetic anisotropy in a material increases the energy required for magnetization reversal in the domains. It opposes the demagnetization field, which causes coercivity in the sample. We believe this enhanced anisotropy resulting from the increased interplanar distance and the associated strain is the key reason for the observation of high coercivity in some of our samples.

Strain-induced magnetism in a 2D material, particularly MoS$_2$, has been reported previously.[13,29,36,47–51] Yun[36] had demonstrated theoretically the change in magneto crystalline energy with strain on a single layer of MoS$_2$. The results showed an increase in Mo-Mo and Mo-S bond length as a function of strain (a) $d_{Mo-S}$ = 2.392, 2.428, and 2.480 Å at 0, 10% and 14% strain (b) $d_{Mo-Mo}$= 2.829, 3.452, and 5.079 Å at 0, 10% and 14% strain. Also, as the strain increased, the spin anisotropy axis switched from the out-of-plane to the in-plane direction.

These intercalation ions could have occupied the sulphur vacancies formed during the 1T phase, so we observe a high coercivity in the samples, even though the saturation magnetization remained lower than previously reported values.[34] Insufficient sulfidation can result in the incorporation of residual Mo–O bonds from the molybdate precursor into the adjacent MoS$_2$ layers and enlargement of the interlayer spacing.[20,42] From the XPS scans of Mo 3p, S 2p, and O 1s, shown in the Supporting Information, the ratio of at% of Mo/S/O is calculated. The Mo/S/O ~ 1:1.63:1.27, 1:1.65:0.99, and 1:1.77:0.37 for S1, S3, and S4, respectively. As the Mo:S ratio increases, the Mo:O ratio decreases. This is consistent with observed decrease in coercivity.

## 3. Conclusions

In this work, the magnetic properties of the 1T phase of MoS$_2$ have been studied at room temperature. The nanosheets were prepared using a single-step hydrothermal synthesis method with a substantial 1T phase. The obtained MoS$_2$ nanosheets exhibit room-temperature ferromagnetism with one of the highest reported coercivities of ~0.3 T. We also see variability in our samples' saturation magnetization and coercivity, which were nominally prepared identically. While these changes occur unintentionally in our samples, our results suggest a strain-driven anisotropy as the possible mechanism for enhanced coercivity. An unequivocally clear correlation between the structural and magnetic characteristics via interplanar distance and magnetic anisotropy reveals the effect of strain caused by intercalation ions in the $MoS_2$ lattice, the latter performing the dual role of lattice enlargement and stabilization as well as modifying the magnetic anisotropy and enhancing the coercivity. Our results present an experimental way to tune the coercivity in a 2D material over two orders of magnitude, which can be useful for applications in various areas, from spintronics and quantum-technology-based devices with lower energy consumption to the development of a new generation of devices employing ferromagnetism in TMDC materials. The results give motivation for a more controlled future study where the interplanar distance can be tuned more carefully using the intercalation process.

## 4. Experimental Section

### 4.1 Measurement Details

The synthesized samples are structurally and magnetically characterized using XRD, Raman, XPS, VSM, HRTEM and EPR. Raman spectra were obtained using a Horiba Jobin Yvon HR800 Raman spectrometer with a 488 nm laser as the excitation source, 1800 lines/mm grating with a resolution of 2 $cm^{-1}$, one mW optical power on the sample, and an integration time of 20 s. High-resolution TEM (HRTEM) analyses were performed on a JEOL 3010 with a Gatan 794 CCD camera. Electron paramagnetic resonance (EPR) was measured using a JES FA200 continuous-wave electron spin resonance spectrometer at an X-band frequency (9.446 GHz) under one mW microwave power. X-ray photoelectron spectroscopy (XPS) analysis was performed on a PHI5000 Versa Probe III spectrometer. M-H measurements were carried out at 300 K using a PPMS-Dynacool with VSM from Quantum Design, USA.

### 4.2 Synthesis details

Ammonium Molybdate Tetrahydrate [$(NH_4)_6Mo_7O_{24} \cdot 4H_2O$] and Thiourea [$NH_2CSNH_2$] were used as Mo and S precursors as shown in Figure 1(b). Stoichiometric amounts of the precursors in the mole ratio 1:2.14 were put in DI water and stirred for 2 hours until fully dissolved. The solution was transferred to a Teflon-lined stainless-steel autoclave kept in a muffle furnace. The reaction temperature was set to 200 °C for 24 hours. The final product was collected after cleaning in DI water and ethanol through centrifugation, followed by vacuum drying in an oven at 60 °C for 6 hours. The dried $MoS_2$ nanosheets are then collected. Six $MoS_2$ nanosheets were synthesized using this single-step hydrothermal method.

### 5. Supporting Information

Supporting Information is available from the author upon request.

### 6. Acknowledgements

We thank ARCI, IIT Madras Research Park for their PPMS facility; SAIF, IIT Madras for their EPR facility. The author acknowledges Prof M. S. Ramachandra Rao for the XPS system at MSRC, IIT Madras. P. K. N. acknowledges the funding from Anusandhan National Research Foundation (ANRF), Government of India, under grant CRG/2023/002425 and Quantum Research Park, IISc Bengaluru, India with grant no QuRp, FSID/2024-25/QP/03. V. P. B. acknowledges the startup grant IP21221797PHNFSC008931 from IIT Madras, India.